\documentclass[aps,prb,twocolumn,groupedaddress,showpacs,showkeys]{revtex4}
\usepackage{graphicx}

\input{tcilatex}
\begin{document}

\title{Anderson localization and the Planck length as source of disorder }
\author{J. C. Flores$^{a}$ and M. Bologna$^{b}$}
\affiliation{(a) Instituto de Alta Investigaci\'{o}n IAI, Universidad de Tarapac\'{a},
Casilla 7-D Arica, Chile.}

\begin{abstract}
The role of disorder on wave propagation through the universe is studied.
Assuming space fluctuations of the order of the Planck length and the size
of the universe as the corresponding localization length for the background
radiation, we obtain the exponent $\upsilon $ (close to unity) in the power
law relationship between these quantities. This suggests that the role of
Anderson localization is not negligible at cosmological scales.
\end{abstract}

\maketitle

\address{ (b) Center For Nonlinear Sciences, University of North Texas, \
P.O. Box 311427, TX 76203, USA}

\address{(a) Instituto de Alta Investigaci\'{o}n IAI, Universidad de Tarapac\'{a}, Casilla 7-D Arica, Chile.\\
(b) Center For Nonlinear Sciences, University of North Texas, P.O. Box
311427, TX 76203, USA}

\section{Introduction}

Anderson localization is a fruitful topic embracing a multitude of wave
phenomena going from quantum to classical physics. This is natural since it
is related to interference of generic coherent waves. In quantum physics,
more explicitly in electronic systems, it has played a major role in our
knowledge of mesoscopic \cite{imry},\cite{brandes} and modern aspects of
nanosystems \cite{brandes}. In classical physics, it is related to a broad
spectrum of situations including electromagnetic-waves, mechanic-waves \cite%
{sheng}, seismic-waves \cite{hennino,bran} and also to the interface theory
called quantum chaos \cite{haake}. In this work we consider another
application of Anderson localization: its possible role in wave propagation
inside the universe. Today it is usually assumed that the average mass
density in space is about $\sim 10^{-28}$[kg/m$^{3}$] but, as mentioned in
literature, there are obviously matter fluctuations \cite{foster,weinberg},
space-time fluctuations \cite{lee}, and then disorder.

We will assume that we have a classical disordered medium and the only
relevant length characterizing distances is the position averaged
fluctuation $\Delta R$. Moreover, in average the system is assumed to be
homogenous and time independent. Consider a wave of frequency $\omega $
propagating in the disordered medium characterized by the velocity $c$ when
disorder is not present. That is, we have the wavelength (with a factor $%
2\pi $) $\lambda ^{\ast }=c/\omega $ characterizing the propagating wave.

As said before, in this paper we are concerned with the possible role of
Anderson localization for wave propagation in the universe, a subject not
yet studied to the best of our knowledge. We will derive a simple
relationship between disorder, the wavelength and the so-called localization
length (section 2). This relationship will be tested using cosmological
parameters (section 3).

\section{ Localization \ length for classical waves}

It is well known that uncorrelated disorder \cite{lyra} could produce
exponential localization \cite{enz} for coherent wave propagation \cite%
{gurtvitz}. Namely, the wave-amplitude decays exponentially with a
characteristic length $L_{c}$ called the localization length. This length is
generally dependent on the degree of disorder (here $\Delta R$), the
dimension $D$ of the system and the frequency $\omega $ (or the wavelength)
of the wave. For electronic systems there is a reasonable comprehension of
the relation between localization and the respective dimension $D$ through
the so-called scaling function \cite{brandes,enz}. It is generally accepted
that for low dimension (smaller than 3) localization holds \cite%
{imry,sheng,enz}. Note that eventual dependence on the mean free path is
assumed inside the wavelength.

Classically, waves with long wavelength are less localized. This is
reasonable since the frequency $\omega $ usually multiplies the disorder
(random refraction index). In one dimension there are explicit examples \cite%
{sheng} where $L_{c}\sim \omega ^{-2}$.

Considering the above statements (including section I) we conjecture that
the localization length for classical systems is given by the generic
relationship:

\begin{equation}
\frac{\lambda ^{\ast }}{L_{c}}=g_{D}\left( \frac{\Delta R}{\lambda ^{\ast }}%
\right) ,  \label{loc-length}
\end{equation}%
where $g_{D}(x)$ is an unknown function depending on the effective dimension 
$D$. The fact that in equation (\ref{loc-length}) the dimensionless quantity 
$\lambda ^{\ast }/L_{c}$ \ (or $\Delta R/\lambda ^{\ast }$)) appears
explicitly is easy to understand since we have normalized all distances by
the wavelength in the generic wave equation.

When no-disorder exists ($\Delta R=0$) the localization length must be
infinite then necessarily the function $g_{D}$ satisfies

\begin{equation}
g_{D}\left( 0\right) =0.  \label{g-function}
\end{equation}

It is well known that disorder tends to smooth out some quantities like the
density of states \cite{brandes} and others. In this sense, we will assume
that near-zero the function $g$ has a power law behavior, particularly,

\begin{equation}
g_{D}\left( x\right) \sim x^{\upsilon }\text{, for }x\ll 1.
\label{g-expansion}
\end{equation}

This is the case for some systems where explicitly $\upsilon =1$ (see
reference \cite{sheng}) and others with a particular random metric \cite%
{flores-bologna}.

\section{ Planck length as a positional disorder source}

In this section we consider a simple model where space is treated
quantum-mechanically but propagating waves are not. Fluctuations are within
the realm of quantum mechanics and they are expected to play a fundamental
role at Planck-length scales of the space-time structure \cite%
{lee,straumann,arzano,amelio}. Consider a classical wave of wavelength $%
\lambda ^{\ast }$ propagating inside the universe. Assume that quantum
fluctuations in the space are of the order of the Planck length $\Delta
R\sim 10^{-35}[m]$. On the other hand, from the above considerations
(section II) the localization length, at low frequency, satisfies the
equation

\begin{equation}
\frac{\lambda ^{\ast }}{L_{c}}=\left( \frac{\Delta R}{\lambda ^{\ast }}%
\right) ^{\nu }.  \label{relation}
\end{equation}

Since the background radiation \cite{foster} ($\lambda \sim 10^{-3}[m]$)
fills the space, and it occupies a localized region of the order of the size
of the universe ($L_{c}\sim 10^{26}[m]$) then the exponent \ $\upsilon $
could be evaluated from (\ref{relation}) as:

\begin{equation}
\upsilon =0.955,  \label{nu}
\end{equation}%
close to unity. Note that for a particular theoretical model with a random
metric the value \ $\nu =1$ was found explicitly\cite{flores-bologna}. The
relationship suggested by the expressions (\ref{relation}) (and (\ref{nu}))
correlates parameters like the universe size, the background wavelength and
the Planck length.

We have considered here only spatial disorder (not mass disorder). Namely,
the only relevant quantity characterizing disorder was $\Delta R$. In the
case of general relativity, this is correct since there is an equivalence
between mass and distance (for instance, through the constant $G/c^{2}$ the
Planck mass is defined from the Planck length \cite{straumann}). Note that
here the approximation (\ref{g-expansion}) is valid since $x=\frac{\Delta R}{%
\lambda ^{\ast }}\sim 10^{-32}\ll 1$.

Our calculations are on static disordered medium (section 1) and the
universe is dynamic. So, the expression (\ref{relation}) is only an
approximation. Nevertheless, dynamic aspects become contained inside the
value of $\lambda ^{\ast }$ (background radiation) which is a time depending
quantity.

\section{Conclusions}

In an averaged-homogenous classical medium supporting waves, where the
disorder is characterized by distance fluctuations (only relevant space
length scale), we conjecture that the localization length satisfies the
relation (\ref{loc-length}). For small frequency, where weak localization is
expected, we have the simple power law relationship (\ref{relation}).

Assuming the origin of space fluctuations as a quantum mechanics phenomenon
(order of the Planck length), the background radiation localized at
distances of the size of the universe then we have evaluated the exponent $%
\nu $ (see (\ref{relation})) giving a number close to unity (\ref{nu}). This
suggests that Anderson localization is a phenomenon which must be considered
in cosmological problems. Naturally the problem involves general relativity
and Anderson localization theory, it is a complex study. From this point of
view, this work is essentially destined to put both branches together in a
simple way.

Acknowledgement J. C. Flores was supported by FONDECYT\ project \ 1070597.

\end{document}